\documentclass[%
 amsmath,amssymb,
 reprint,%
]{revtex4-1}

\usepackage{url}
\usepackage{graphicx}
\usepackage{caption}
\usepackage{subcaption}

\usepackage{dcolumn}
\usepackage{bm}
\usepackage[utf8]{inputenc}
\usepackage[T1]{fontenc}
\usepackage{mathptmx}
\usepackage{amsmath}
\usepackage{wrapfig}
\usepackage{physics}
\usepackage{listings}
\usepackage{xcolor}

\begin{document}

\preprint{AIP/123-QED}

\title{Oscillation and collective behaviour in convective flows}

\author{A. Gergely}
\affiliation{Physics Department, Babe\c{s}-Bolyai University, Cluj-Napoca, Romania}  
\author{Cs. Paizs}
\affiliation{Faculty of Chemistry and Chemical Engineering, Babe\c{s}-Bolyai University, Cluj-Napoca, Romania }
\author{R. T\"ot\"os}
\affiliation{Faculty of Chemistry and Chemical Engineering, Babe\c{s}-Bolyai University, Cluj-Napoca, Romania }
\author{Z. N\'eda}
\email{zoltan.neda@ubbcluj.ro}
\affiliation{Physics Department, Babe\c{s}-Bolyai University, Cluj-Napoca, Romania}

\date{\today}

\begin{abstract}
 Oscillation and collective behavior in convection-driven fluid columns are investigated and discussed in analogy with similar phenomenon observed  for the flickering flames of candle bundles. 
It is shown experimentally that an ascending circular Helium gas column performs an oscillation which is similar in several aspects to the oscillation of diffusion flames.  Increasing the nozzle diameter leads to a decrease in the oscillation frequency, while increasing the flow rate results in an increase in this frequency. For helium columns oscillating at nearby frequency and placed close to each other anti-phase synchronization and beating phenomena is observed. A simple toy-model based on elementary hydrodynamics describes the observed oscillations and leads to oscillation frequencies in the right order of magnitude.  

\end{abstract}

\maketitle

\section{Introduction}
    Oscillation and collective behavior of diffusive flames is an intriguing and well studied problem \cite{Chamberlin1948,Durox1995,DUROX1997,Huang1999,Kitahata2009,Ghosh2010,Okamoto2016,Chen2019,Gergely2020}. Experimental results suggests however \cite{Yuan1994}, that similar oscillations are present in rising gas columns, and therefore the two phenomenon could be more strongly related than it is nowadays believed. More specifically one could ask, whether the existence of the flame or the chemical reactions inside of it, captured in the currently used models is a necessary ingredient to understand the oscillations in diffusive flames, or hydrodynamics by itself  is enough to understand this interesting phenomena.  Some carefully conducted experiments and computer simulations could reveal more analogies between the two phenomenon and therefore could help in a better understanding for both of them. The present work intends to contribute in such directions.

 A candle is a simple system consisting of a wick and the surrounding  combustible material (usually paraffin). The fuel of the candle does not burn in solid form, for the combustion to take place we must first evaporate the fuel. The combustion reaction takes place in the boundary layer of the fuel vapor and air, based on which the candle flames are associated with diffusion flames. It has long been known \cite{Chamberlin1948} that under certain conditions the volume of diffusion flames changes periodically over time, this phenomenon is called the oscillation of diffusion flames.

At normal atmospheric oxygen concentration (21\% $O_2$), the flame of a single candle burns in a stable manner. In order to obtain oscillations the candles must be arranged in a bundle. Different physical parameters affects the oscillation frequency of the candle bundle. In our previous work \cite{Gergely2020}, we investigated experimentally how the oscillation frequency changes as a function of the number of candles in the bundle and how the oxygen concentration around the candle bundle affects the oscillation. We have shown that for the compact and hollow arrangements of the candles inside the bundle the oscillation frequency decreases  as the number of candles are increased in the bundle. We also proved that as the oxygen concentration increases, the oscillation frequency decreased. We  observed that a high oxygen concentration can cause oscillation in cases where this would not occur at a normal oxygen concentration. Interestingly, high oxygen concentration can also stop the oscillations in cases where this would occur at a normal oxygen concentration.

If the flame of two candle bundles are placed nearby each other, collective behavior in form of synchronization appears as a result of the interaction between their flickering \cite{Kitahata2009}. We thoroughly examined this collective behaviour as a function of the distance between bundles by a properly defined synchronisation order parameter \cite{NZarticle,Gergely2020}. It was found that for small flame distances, in-phase synchronisation develops. At larger distances this turns sharply in counter-phase synchronization and by further increasing the distance between the bundles one observes a graduate decrease in the synchronization level. 

In the seminal work of Kitahata et. al \cite{Kitahata2009}, the authors conclude that the coupling mechanism is originated in the thermal radiation emitted by the candles. The phenomena of flame oscillation and the synchronization of nearby flames is modeled by two coupled evolution equations in which the temperature and oxygen concentration inside the flame are the relevant dynamical variables. Collective behavior is accounted by considering similar evolution equations for the interacting flames and a coupling mechanism between them through thermal radiation. The experimental results presented in \cite{Gergely2020} contradicted however the existence of such a coupling mechanism and lead to an improved dynamical system model both for the oscillation phenomena and for the observed synchronization. The model proposed in \cite{Gergely2020} is similar with the original model of Kitahata et. al and  it is still based on the chemical reactions that takes place inside the flame. In this improved model the coupling mechanism is realized via the oxygen flow induced by the periodic changes in the flame size. Interestingly, this improved and oversimplified model described excellently all the carefully gathered experimental results. 

Intriguing similarities with some purely hydrodynamical phenomena could seriously question however whether the existence of a flame or a chemical reaction inside of it is needed in order to understand the oscillation and synchronization phenomena. For example, in \cite{Yuan1994} , the authors observed that a Helium column flowing laterally into the air performs similar oscillations to the one in diffusion flames. This raises the possibility that the role of flame in our candle experiments is only to create the convective flow in which hydrodynamic instabilities occur and this flow is causing all the interesting physics connected to the oscillation of diffusive flames.  In the followings we will approach this question both experimentally and theoretically.  We present experimental results pointing to quantitative analogies. On the theoretical side, we consider a simple analytical approach based on elementary hydrodynamics which is intended to estimate the oscillation frequencies as a function of the relevant parameters of the flow. Finally, we discuss all the qualitative and quantitative analogies revealed between the dynamical behavior of diffusion flames and the ones observed in convective flows.      

\section{Experimental studies}
\subsection{Experimental setup}

 With the experiments detailed below, we are looking to answer whether hydrodynamic instabilities in convective flows are able to explain by itself the oscillations and collective behavior observed for diffusive flames.

The flow must be produced without any chemical reaction, therefore a controlled flow of Helium gas in air was considered to be a good solution. Since the flow of Helium in air does not emit visible light,  the experimental methods used for candle flames \cite{Gergely2020} had to be modified. The Schlieren technique \cite{settles2001schlieren,Leptuch2006} proved to be a simple and efficient method to visualize the flow in rising Helium columns.

\begin{figure}
\centering 
\includegraphics[width=.45\textwidth]{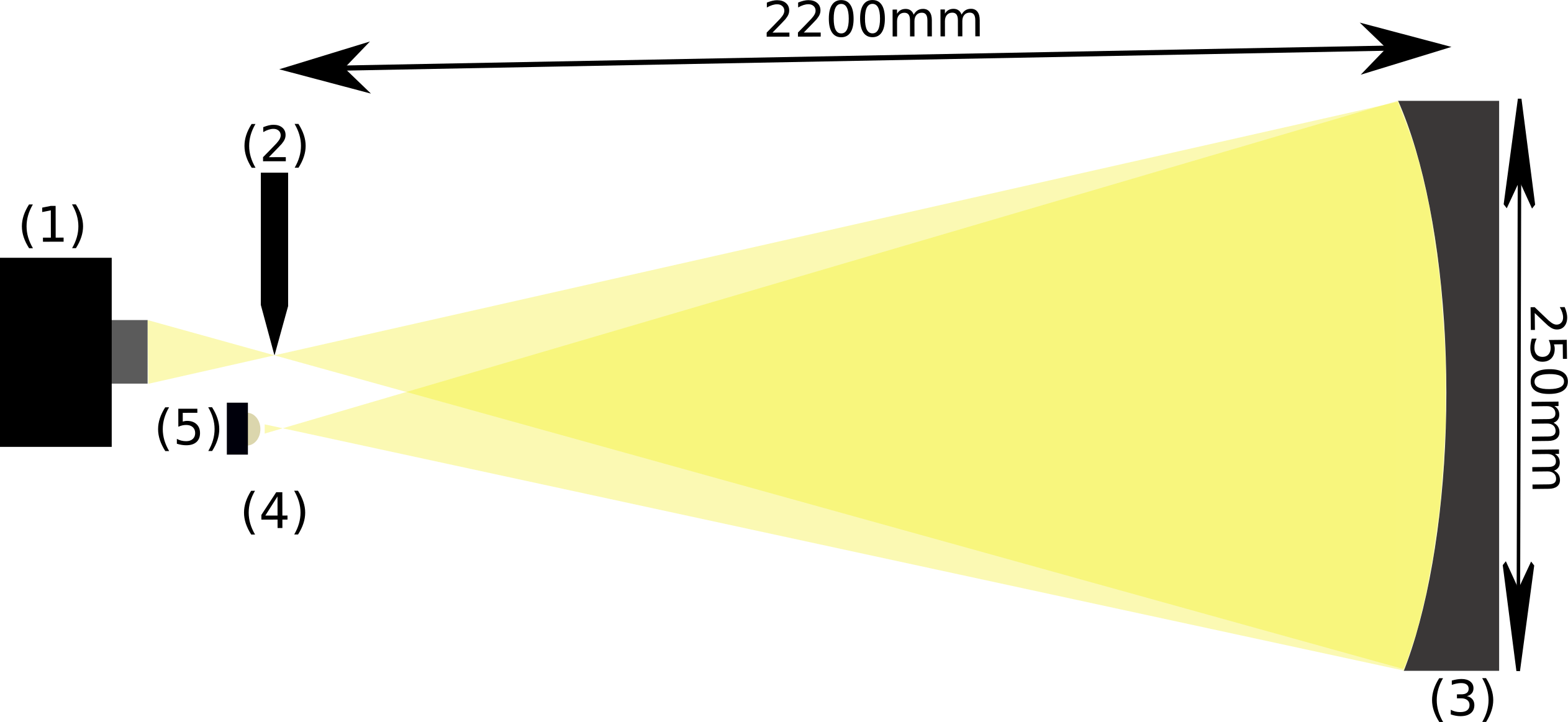}
\caption{Schematic drawing of the used Schlieren setup. The following elements are indicated: (1) -fast camera, (2) -razor blade, (3) -parabolic mirror, (4) -circular gap, (5) -light-emitting diode.}
\label{fig:Schlhliren}
\end{figure}

Schlieren imaging is a method for visualising and qualitatively characterising the refractive index distribution in  a transparent medium. There are several types of Schlieren equipments, we have built one of the cheapest and simplest off-axis mirrored version presented in \cite{settles2001schlieren} page 47. A schematic diagram of our equipment is shown in Figure \ref{fig:Schlhliren}.

The operation principle is relatively simple. With the help of a light-emitting diode, a circular gap is illuminated, creating a nearly point-like light source. This light source is imaged using a parabolic mirror, and a blade is placed at the location of the image so that it obscures half of it. The uncovered light enters the camera lens used for recording. If there is anisotropy in the refractive index in the space between the light source and the mirror, a refraction will occur, resulting a change in the amount of light obscured by the blade. The change in brightness results in dark and light areas in the camera image, making the refractive index gradients visible. The refractive index of Helium used in our experiments is lower than the refractive index of air at room temperature, so their flow can be observed using this technique.
The largest change in brightness are in those places where the greatest amount of light refraction due to optical anisotropy occurs. The anisotropy results in an increase or decrease in brightness depending on the direction of light refraction and the location of the blade in the equipment. For example
Figure \ref{fig:SchlirenExample} shows frames made by the Schlieren technique on a  rising Helium  column. In these cases an increase in brightness relative to the background occurs in those parts of the images where the thickness of the optically less dense Helium decreases from left to right. In the case of the projection seen by the camera, the largest change in the layer thickness of the Helium is at the edges of the column, so the largest increase in brightness will occur there. Due to the high-frequency oscillations observed in the gas columns (above 10 Hz), for recording we used the Sony Cyber-Shot DSC-RX100 VI photo camera which allowed HD format recording at 250 fps. 
 
In our experiments we monitored the movement of these edges at a given height.  For the proper image processing algorithm the Otsu method for the gray-scale images was selected \cite{Otsu-original,Otsu:2020}.  This method consists of choosing a critical intensity level below which the value of the recorded pixel is assigned zero, and otherwise it is assigned the logical one. If the Otsu threshold is selected properly, the pixels near the boundary layer will be 1 and the other pixels will become zero. After applying the 
cutoff, going from left to right we identify the first pixel whose value is one, approximating the location of the boundary layer at a given height. Movies with original recordings and the ones processed with the Otsu method can be consulted on our YouTube channel \cite{Attila3}. 

\begin{figure}
\centering 
\includegraphics[width=.45\textwidth]{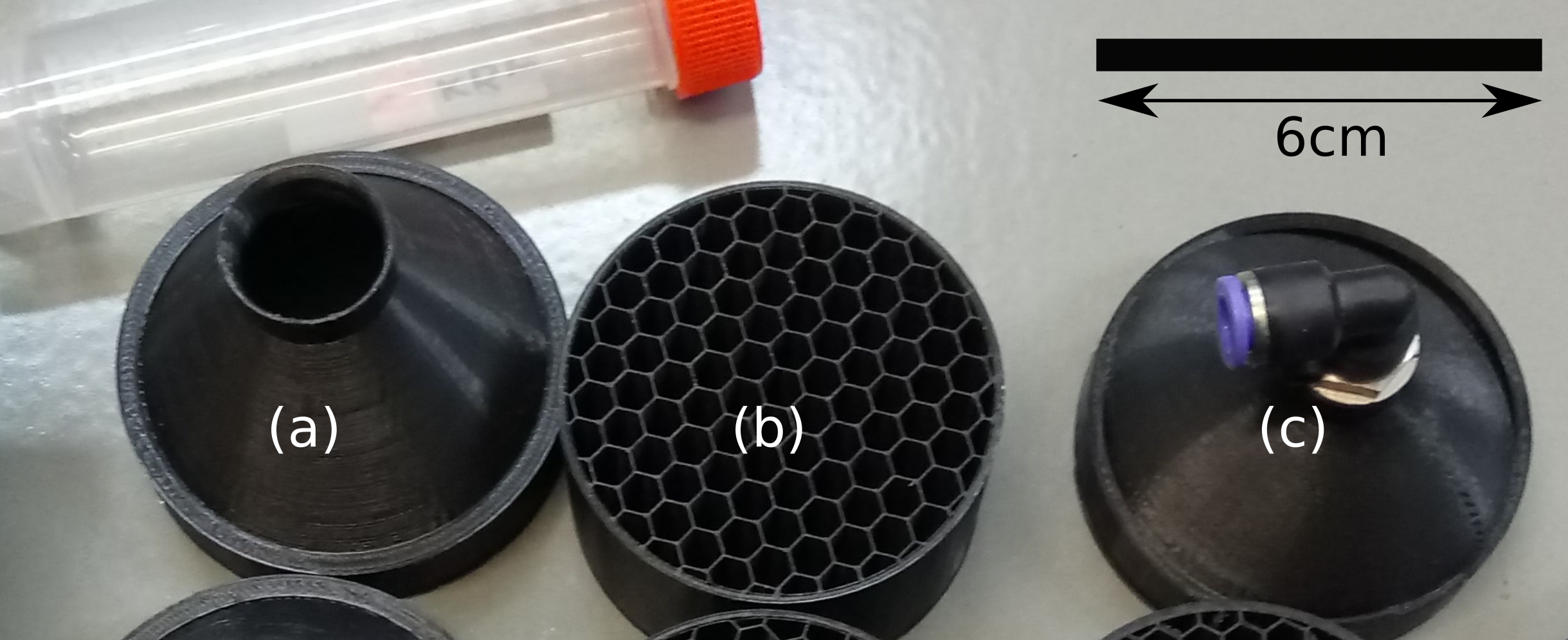} 
\caption{Elements of the nozzles realised with a 3D printer. Element (a) is the outlet of the nozzle, element (b) is the hexagonal structure that ensures the laminar nature of the flow, and element (c) is the inlet through which Helium is introduced into the nozzle.}
\label{fig:Schliren3Dprinted}
\end{figure}

\begin{figure}
\centering 
\includegraphics[width=.45\textwidth]{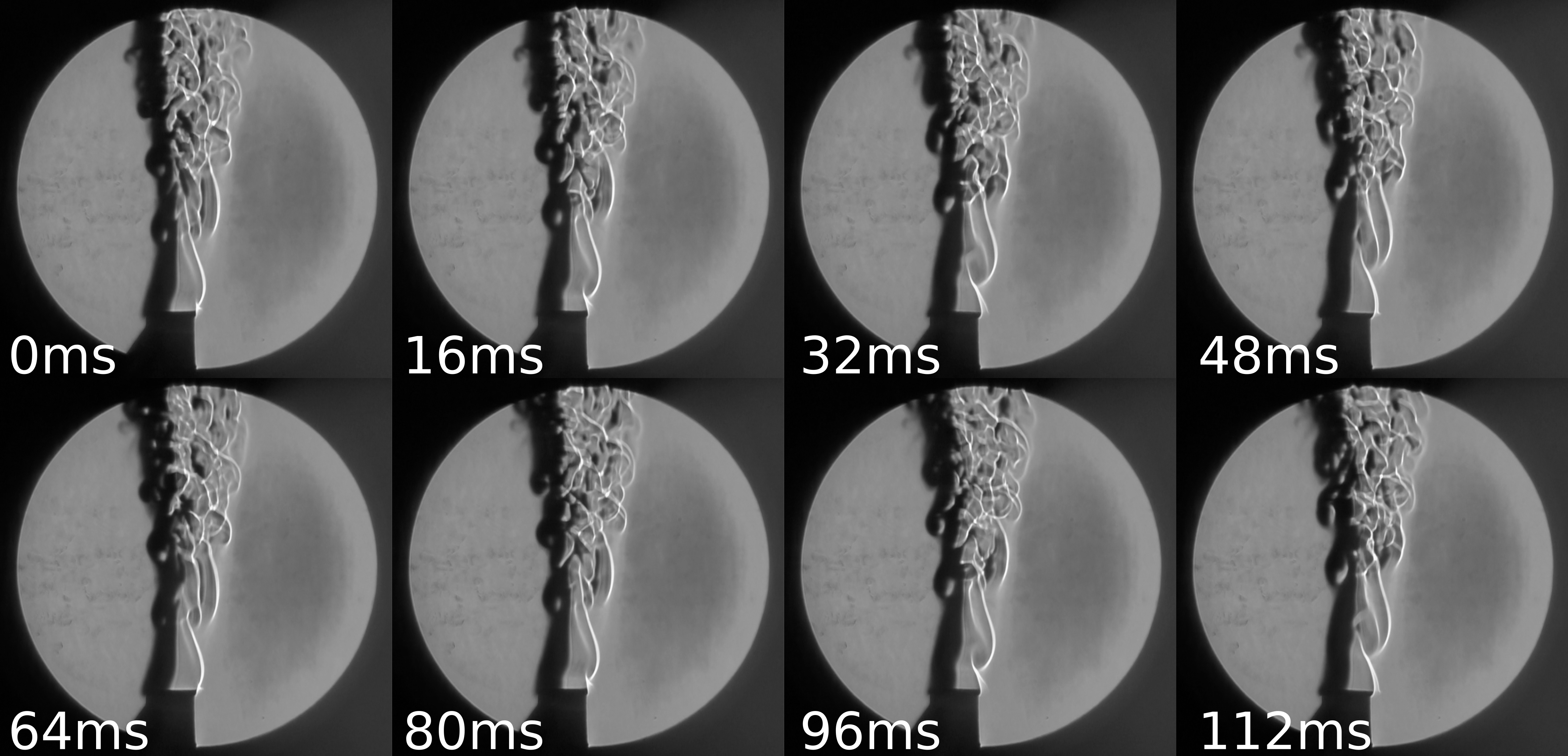} 
\caption{Time-lapse images taken using the Schlieren technique for a column of Helium gas flowing out in vertical direction from the nozzle.} 
\label{fig:SchlirenExample}
\end{figure}

The nozzles used to initiate the Helium column were realized by 3D printing. They consist of 3 parts: an inlet part, a hexagonal structure that ensures the laminar nature of the initial flow, and an outlet part through where the gas leaves. The modular design was needed so that the hexagonal structure and the inlet parts did not have to be manufactured separately for the outlets with different diameter, saving by this time and plastic.
Several different plastics (PC, PLA, CPE) and nozzle diameters for the 3D printer were tested (0.25, 0.4, 0.6, 0.8 mm) and the best results were obtained with a combination of 0.4 mm nozzle diameter and PLA plastic. The 3D printed elements are shown in Figure \ref{fig:Schliren3Dprinted}.

Helium was introduced into the nozzle through a $1/4$ inch tube. One defining parameter of the Helium's flow is the debit (yield) of the gas flowing through the nozzle. This was controlled by a needle valve and measured with a precise rotameter.

Time-lapse images obtained with the Schlieren technique are illustrated in 
Figure \ref{fig:SchlirenExample}, and some recorded movies can be consulted in \cite{Attila2}.

\subsection{Results}
\subsubsection{Oscillations}

For studying the oscillation of the ascending Helium column, the jet was  produced with circular cross-section nozzles of various diameters.  Our experiments were performed at room temperature and normal atmospheric pressure, the purity of the used Helium was 99\%. Each experiment was repeated 5 times, and the standard deviation for the obtained results was visualized with error bars on the plots. The video recordings obtained from the experiments can be viewed on the \cite{Attila1,Attila2} youtube playlists.

At a constant Helium yield of $\Phi=$46 $\pm$ 2.3 $l / min$, it was examined the oscillation frequency variation with the nozzle diameter. As shown in Figure 4, it was observed that the increasing nozzle diameter determined the decrease of the oscillation frequency, which can be well approximated by a power law function with exponent $-1.64$. These results are somehow in accordance with those obtained with candle bundles, for which a power-law like trend was also observed for the oscillation frequency as a function of the number of candles in the bundle \cite{Gergely2020}. 
\begin{figure}
\centering 
\includegraphics[width=.495\textwidth]{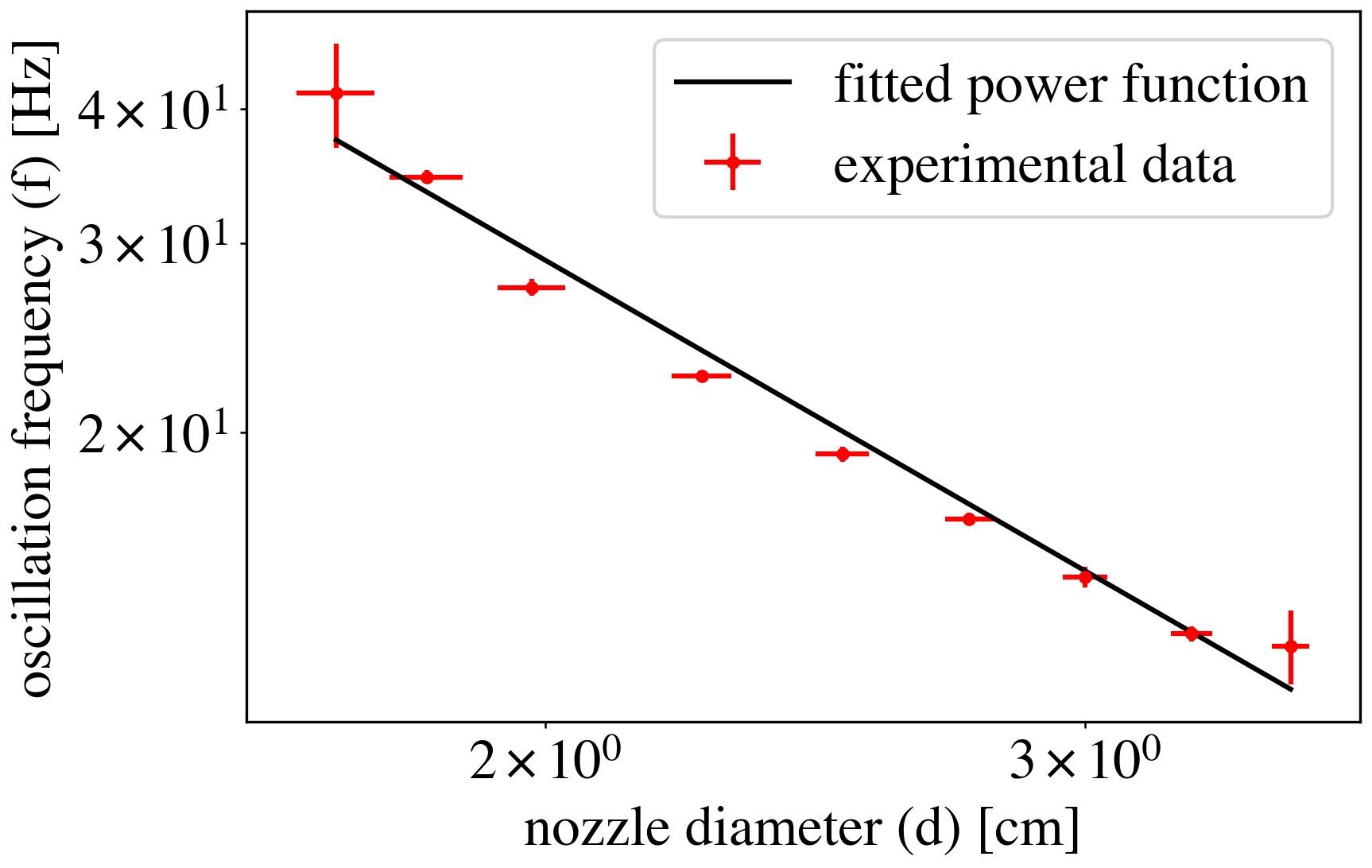}
\caption{Oscillation frequency of the Helium column as a function of the nozzle diameter 
for a yield of $\Phi=$46 $\pm$ 2.3 $l / min$.  The fitted power function is shown with a continuous line.  Please note the double logarithmic scale. (In the case of nozzle diameter, the high margin of error is due to the eccentricity of the nozzle and the error was considered as the difference between the largest and the smallest measured diameter.)}
\label{fig:Oscillation frequency}
\end{figure}

\begin{figure}
\centering 
\includegraphics[width=.495\textwidth]{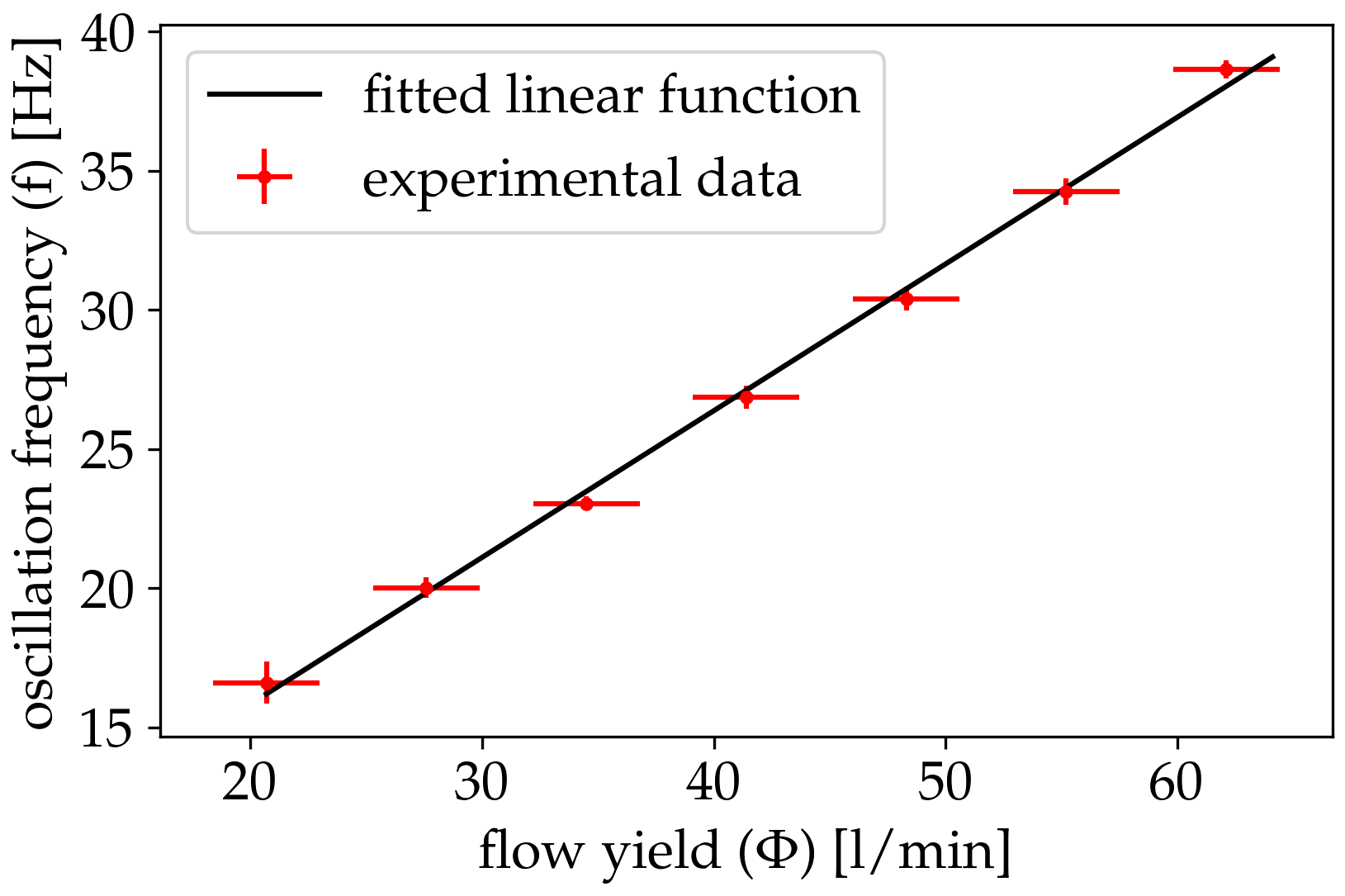}
\caption{Oscillation frequency of the helium column as a function of the yield (flow debit) obtained for nozzle with a diameter of 2 cm. With increasing flow yield, the frequency of the oscillation increases in an almost linear manner. (For the yield, the error was considered to be equal to the flow debit corresponding to one division on the scale of the rotameter.)}
\label{fig:heliumFreqYield}
\end{figure}

The effect of the flow rate was examined using a 2 cm diameter circular nozzle. Our results  are shown in Figure \ref{fig:heliumFreqYield}, where one will observe that as the yield increases, the oscillation frequency increases.  This increase is well approximated by a linear trend.

\subsubsection{Collective behaviour}

For studying the collective behavior of the flow oscillations, experiments with two Helium columns were performed. 

\begin{figure}
     \centering
     \begin{subfigure}[b]{0.45\textwidth}
         \centering
         \includegraphics[width=\textwidth]{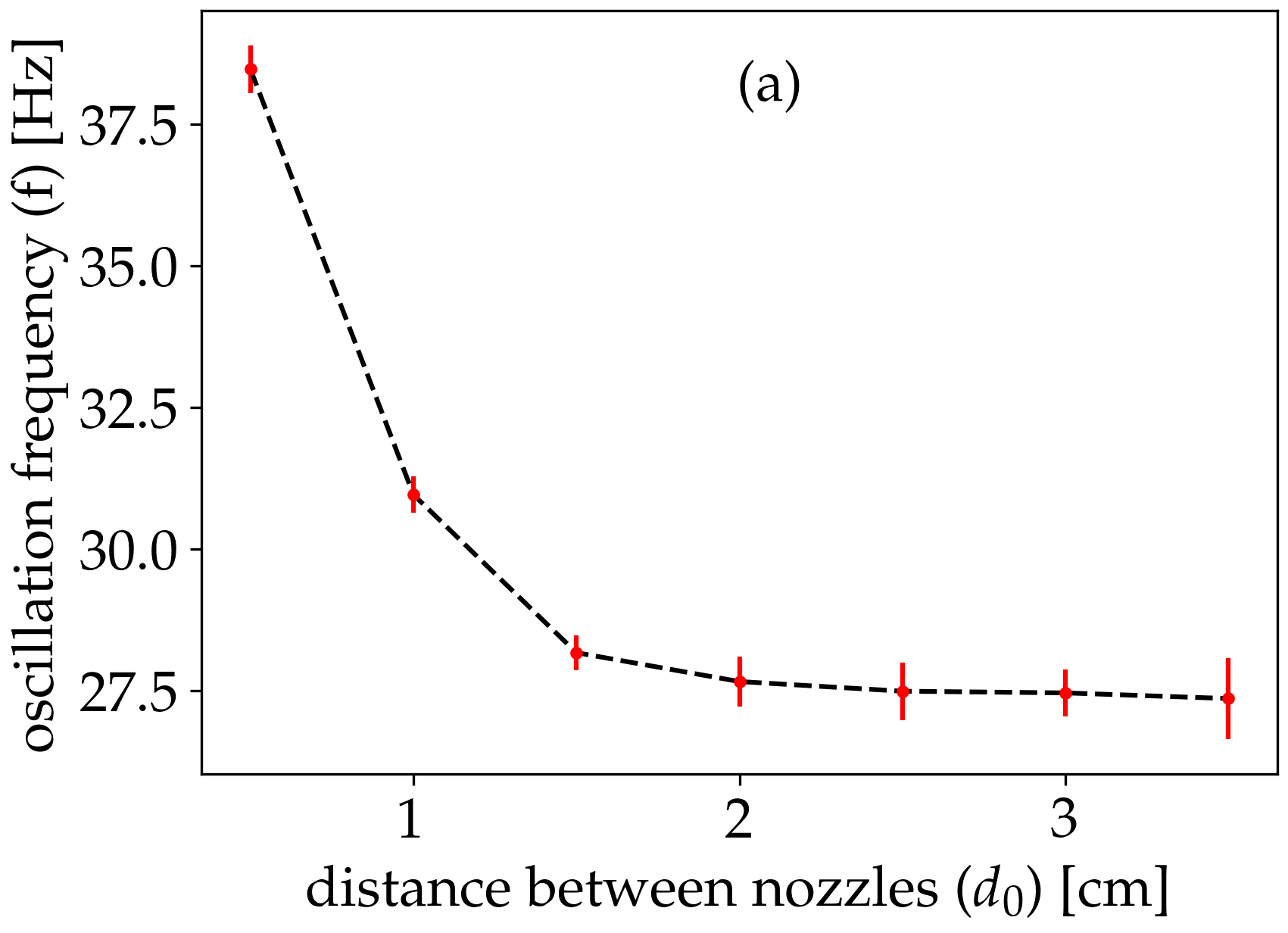}
         \label{fig:y equals x}
     \end{subfigure}
     \begin{subfigure}[b]{.45\textwidth}
         \centering
         \includegraphics[width=\textwidth]{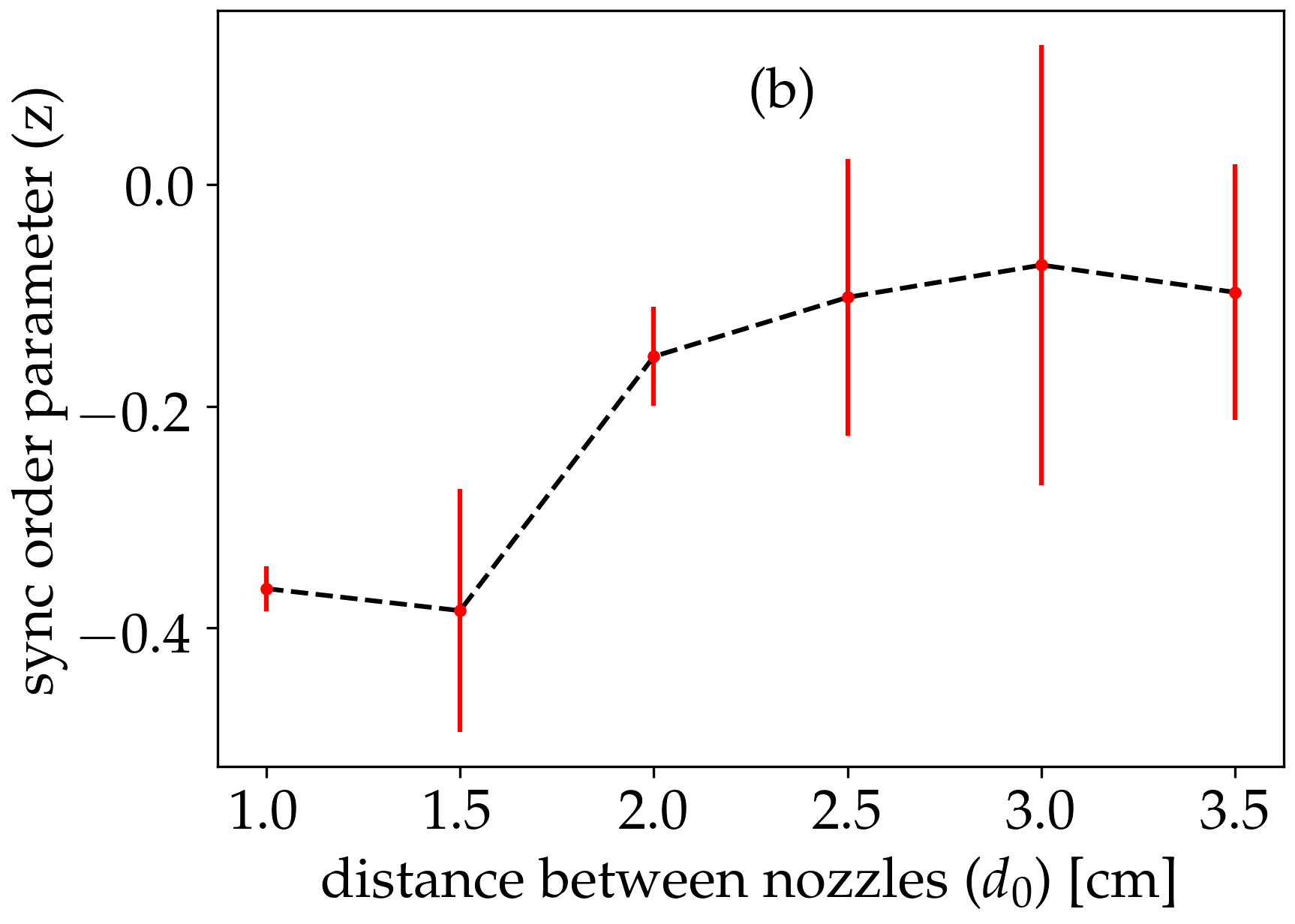}         
         \label{fig:three sin x}
     \end{subfigure}
        \caption{Figure (a) shows the common oscillation frequency of the Helium columns flowing out of two identical 2 cm diameter nozzles with a flow rate of $\Phi$=46 $\pm$ 2.3 l/min as a function of the distance between the nozzles. Figure (b) plots the synchronization order parameter $z$ (defined in \cite{Gergely2020}) obtained from the oscillating time series of the Helium flows.}
      	\label{fig:yield46}
\end{figure}

If two diffusion flames are placed nearby each other  collective behaviour in form of synchronization of the flickering can appear. \cite{Kitahata2009,Forrester2015,Okamoto2016,Manoj2018,Yang2019,Dange2019,Fujisawa2020,Gergely2020}. Similarly with the studies performed on candle flames, first we will examine the collective 
oscillation frequency and the synchronization order parameter $z$ for  two Helium columns with the same flow parameters (yield and nozzle diameter) as a function of the distance between the nozzles. 
The experimental apparatus is pretty much the same as the one used for one Helium column, the only difference is that now two nozzles are used, with a fine control on the distance between their close edges.  Our results are summarized in Figure \ref{fig:yield46}. On Figure \ref{fig:yield46}a we plot the measured oscillation frequency as a function of the distance between the nozzles and in Figure \ref{fig:yield46}b we show the value of the synchronization order-parameter as a function of the nozzles distance.  Results on both graphs are for nozzles of 2 cm diameter and Helium flow rate of 46 $\pm$ 2.3 l / min. The synchronization order-parameter $z$ is a number in the interval $[-1,1]$, an it is defined and determined from the Otsu processed images in the same manner as it was done in \cite{Gergely2020}. The value $z=1$ means complete in-phase synchrony while $z=-1$ indicates a perfect counter-phase synchronization.

From these results and the ones plotted in Figures \ref{fig:Oscillation frequency} and 
\ref{fig:heliumFreqYield} one can conclude that for short distances the frequency is significantly higher than the one observed for non-interacting Helium columns with the same parameters (flow rate and nozzle diameter). It can also be observed that for the entire distance range that we have examined anti-phase oscillation dominates. This is different from the case of the collective behavior observed for candle flames, where at short separation distances in-phase synchronization is also observed. More on such similarities and differences will be discussed in the concluding section. 

As previously observed, another interesting collective behavior for interacting diffusion flames with slightly different frequencies also occurs \cite{Chen2019}. In such case a phenomenon similar to the "beating" known in acoustics is observable. For flickering candle flames, this means that the amplitude of oscillation for one of the flames performs a long period modulation.

In order to test whether such beating is observable for Helium columns as well we use two approaches to produce slightly different oscillating frequencies. In the first approach we kept the Helium yield constant and varied the nozzle diameter, while in the second approach we modified the Helium yield for a constant nozzle diameter.

\begin{figure}
     \centering
     \begin{subfigure}[b]{0.45\textwidth}
         \centering
         \includegraphics[width=1\textwidth]{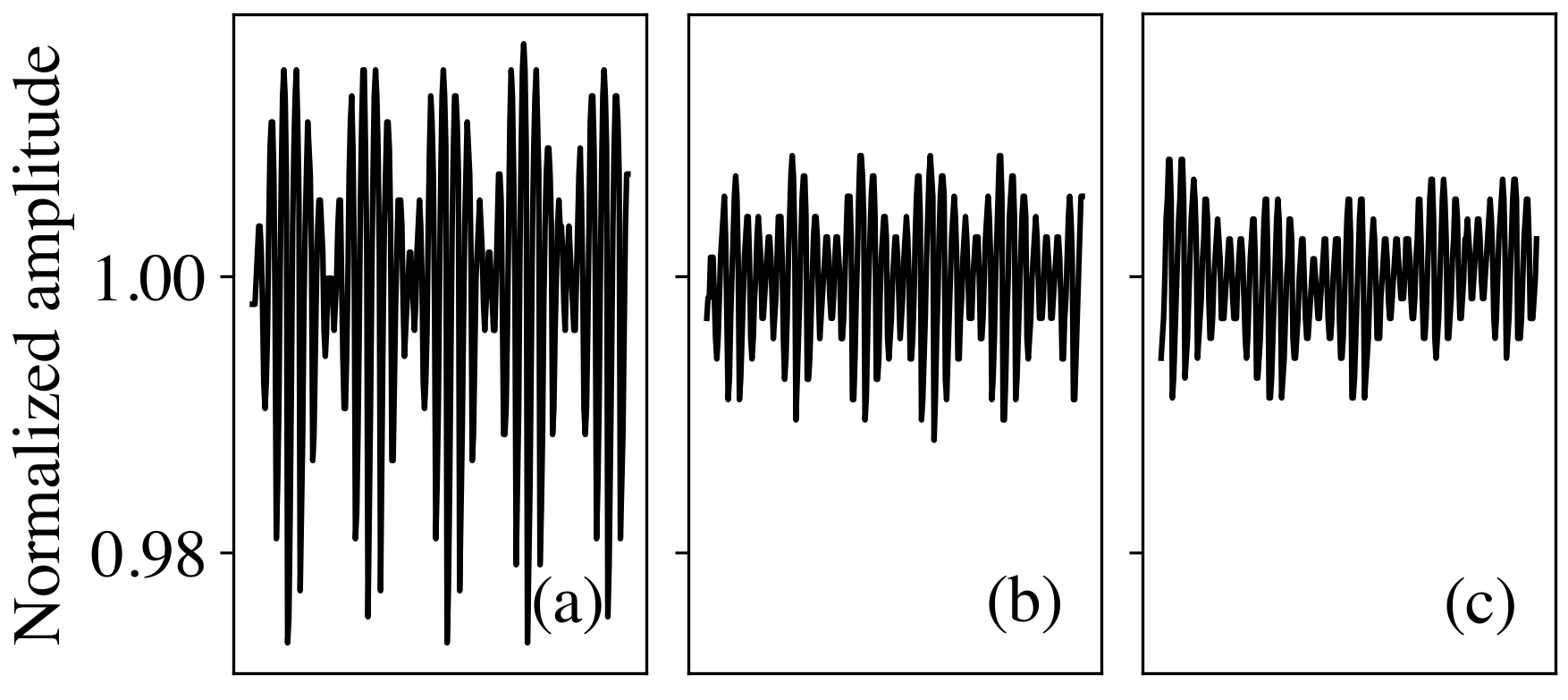}
          \label{fig:y equals x}
     \end{subfigure}
     \hfill
     \begin{subfigure}[b]{0.45\textwidth}
         \centering
         \includegraphics[width=1\textwidth]{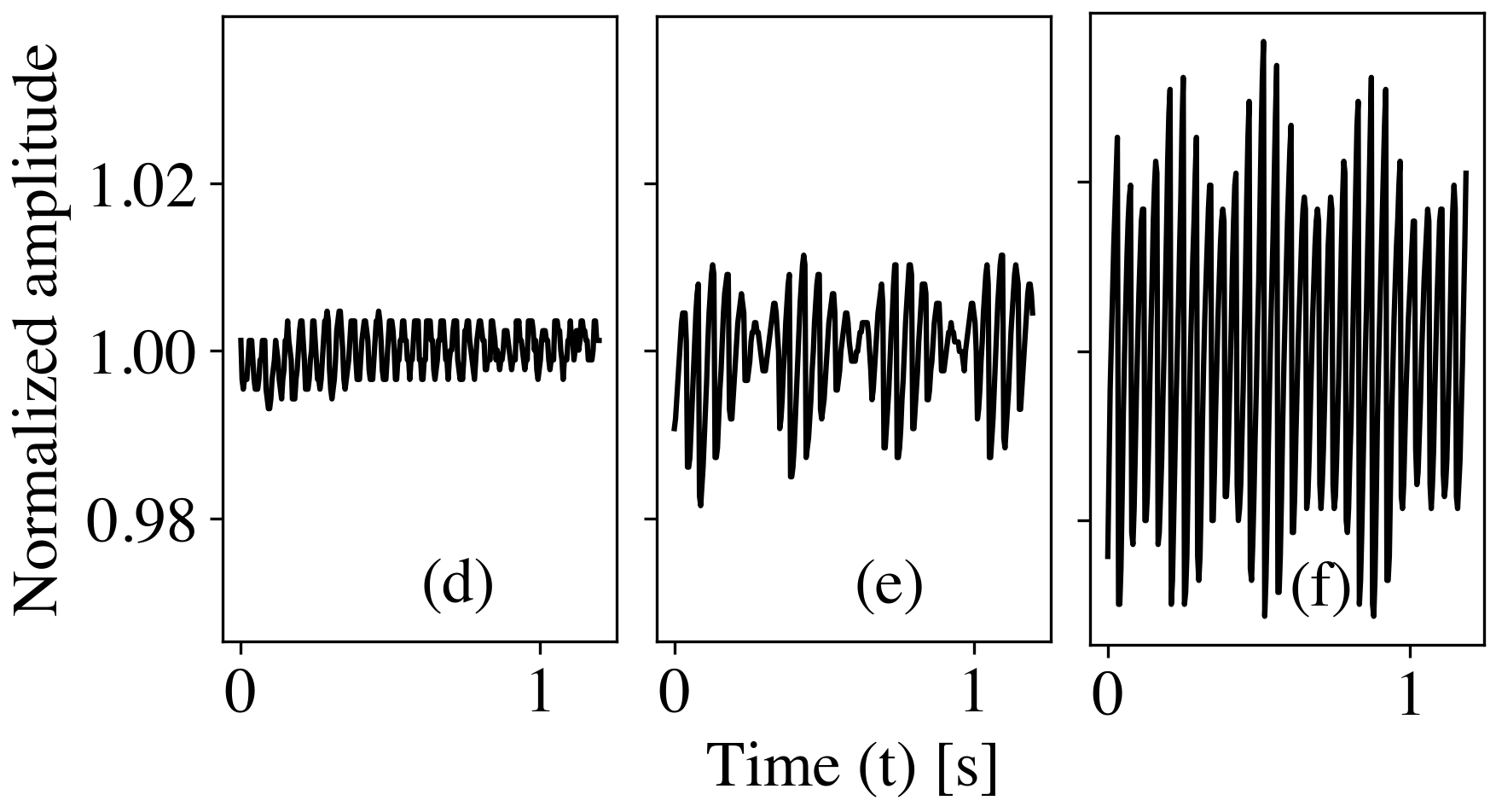}
         \label{fig:three sin x}
     \end{subfigure}
        \caption{Beating-like phenomenon observed for two interacting Helium gas columns. For Figures a, b, c, the distance between the nozzles is 1, 1.5, 2 cm, respectively, and the yield of Helium flow is $\Phi=34.5 \pm$ 2.3 l / min and $\Phi=46\pm$ 2.3 l / min, respectively. The nozzle diameter is 2 cm and the plotted time series are for the oscillation of the Helium column with the lower debit. For Figures d, e, f, the nozzle distances are 1, 1.5, 2 cm, respectively, the Helium flow debit for both nozzles is $\Phi=46 \pm 2.3$ l / min and the nozzle diameters are 2.25 cm and 2.5 cm. The plotted time-series are for the Helium column initiated from the 2.5 cm diameter nozzle.}
      	\label{fig:beating}
\end{figure}

As the graphs in Figure \ref{fig:beating} illustrates we were able to obtain the beating-like phenomenon with both methods. Figure \ref{fig:beating}a,b,c shows the beating realized with different yields. In these experiments the nozzle diameters are fixed for  2 cm  and the outflow yields are $\Phi=$34.5 $\pm$ 2.3 l/min and $\Phi=$46 $\pm$ 2.3 l/min. In these figures the time series of the Helium column with the lower yield is plotted for the distance between nozzles fixed at 1, 1.5 and 2 cm, respectively. 

For the beating phenomena observed with nozzles of different diameters and the same flow yields the time-series are shown in Figure \ref{fig:beating}d,e,f. In these experiments the flow yield is $\Phi=46 \pm$ 2.3 l / min, and the nozzle diameters are 2.25 and 2.5 cm. The plotted time series for the oscillation of the Helium column are for the flow from the larger diameter nozzle. In Figures \ref{fig:beating}d, e, f  the distance between the nozzles are again 1, 1.5 and 2 cm, respectively.

\section{Analytical approach for the oscillation frequency}

We present here a toy-model for understanding the oscillation frequency of the rising Helium gas column as a function of the flow-rate and nozzle diameter. This oversimplified model is unfortunately not appropriate to tackle the problems related to collective behavior.

Our basic assumption is that the Helium column accelerated by gravity becomes unstable at a given Reynolds number. Accepting this, we assume that for a given volume element the time from the outflow up to the formation of instability will approximate the oscillation period of the Helium gas column.
Definitely, this is an oversimplified approach, since the dynamical feedback of this instability on the outflow from the nozzle is neglected and therefore the periodicity for the formation of these instabilities 
are also lost. Nevertheless, we hope that this approach will estimate a correct time-scale for the formation of the first instability and assume that this time-scale drives the periodicity of the observed oscillations.

The Reynolds number for a cylindrical fluid column can be given by the following equation
\begin{equation}
	\begin{split}
		R_e(t)=\frac{2\cdot v(t)\cdot r(t)}{\nu},
	\end{split}
	\label{eq:ray}
\end{equation}
where $v(t)$ denotes the velocity of the considered fluid element, $r(t)$ denotes its radius and $\nu$ denotes the dynamic viscosity of the fluid.

According to our assumption the oscillation period $\tau$ will be estimated as the time necessary for 
the considered cylindrical fluid element's Reynolds number to reach a critical value $R_{e}^c$:
\begin{equation}
	\begin{split}
		R_e(\tau)=R_{e}^c
	\end{split}
	\label{eq:reyc}
\end{equation}

In the following we examine the above model in two cases where simple analytical results can be drawn. In the first limit we neglect viscosity, while in the second approximation 
friction effects are taken into consideration to describe the rising dynamics of the  Helium gas. 
In the later approach we discuss again two cases. First we assume no slip condition for the  air-Helium boundary layer, fixing the velocity of Helium on this interface at 0.  Then we discuss a much more realistic case, where we allow the movement of the air-Helium boundary layer implementing a semi-slip boundary condition.

\subsection{Dynamics with no friction}\label{ch:Setups}
The buoyancy force acting on a Helium element with volume $V$ can be written as:
\begin{equation}
	\begin{split}
		F_V=V\cdot g\cdot (\rho_{Air}-\rho_{He})
	\end{split}
\end{equation}
Here  $g$ denotes the gravitational acceleration, $\rho_{He}$ and $\rho_{Air}$ denote the density of Helium and air, respectively. If one neglects the friction in the air-Helium boundary layer, based on the Newtonian equation of motion, the velocity of a Helium gas element with an initial velocity $v_0$  will be:
\begin{equation}
	\begin{split}
		v(t)=v_0+\frac{ g\cdot (\rho_{Air}-\rho_{He})}{ \rho_{He}}\cdot t
	\end{split}
	\label{eq:veloc}
\end{equation}
To calculate the Reynolds number, we also need the radius $r(t)$ of the cylindrical Helium column element at time-moment $t$. This is determined from the continuity equation as follows:
\begin{equation}
	\begin{split}
		r(t)=r_0\cdot\sqrt{\frac{v_0}{v(t)}}
	\end{split}
	\label{eq:radius}
\end{equation}
We denoted by $r_0$ and $v_0$ the radius and the velocity of the Helium gas column at the moment of outflow from the nozzle, respectively. From equations (\ref{eq:radius}), (\ref{eq:veloc}) and (\ref{eq:ray})  one gets the Reynolds number for the flow at time $t$:
\begin{equation}
	\begin{split}
		R_e(t)=\frac{2\cdot r_0\cdot \sqrt{\left(v_0+\frac{ g\cdot (\rho_{Air}-\rho_{He})}{ \rho_{He}}\cdot t\right)\cdot v_0}}{\nu}
	\end{split}
	\label{eq:rey1}
\end{equation}
Using this result and our approximation (\ref{eq:reyc}) for the estimation of the $f=1/\tau $ oscillation frequency we get: 
\begin{equation}
	\begin{split}
		f=\frac{ g\cdot (\rho_{Air}-\rho_{He})\cdot v_0\cdot r_0^2}{\left({\left(\frac{R_e^c\cdot \nu}{2}\right)}^2-{(v_0\cdot r_0)}^2\right)\cdot  \rho_{He}}
	\end{split}
	\label{eq:rey2}
\end{equation}
In the above equation, the values of all parameters are known from the experiments except for the critical Reynolds number $R_{e}^c$. 
Using  a realistic estimate for the critical Reynolds number: one between a laminar and turbulent flow, the model gives a correct order of magnitude for the oscillation frequency and also correctly reproduces the experimentally obtained trends for oscillation frequency as a function of nozzle diameter and outflow yield.
Reconsidering now  equation (\ref{eq:rey2}) to include that flow yield $\Phi$ instead of flow velocity $v$ we get: 

\begin{equation}
	\begin{split}
		f=\frac{ g\cdot \left(\rho_{Air}-\rho_{He}\right)\cdot \frac{\Phi}{\pi}}{\left({\left(\frac{R_{e}^c\cdot \nu}{2}\right)}^2-{\left(\frac{\Phi}{r_0\cdot\pi}\right)}^2\right)\cdot  \rho_{He}}
	\end{split}
	\label{eq:rey3}
\end{equation}

\begin{figure}
         \includegraphics[width=0.45\textwidth]{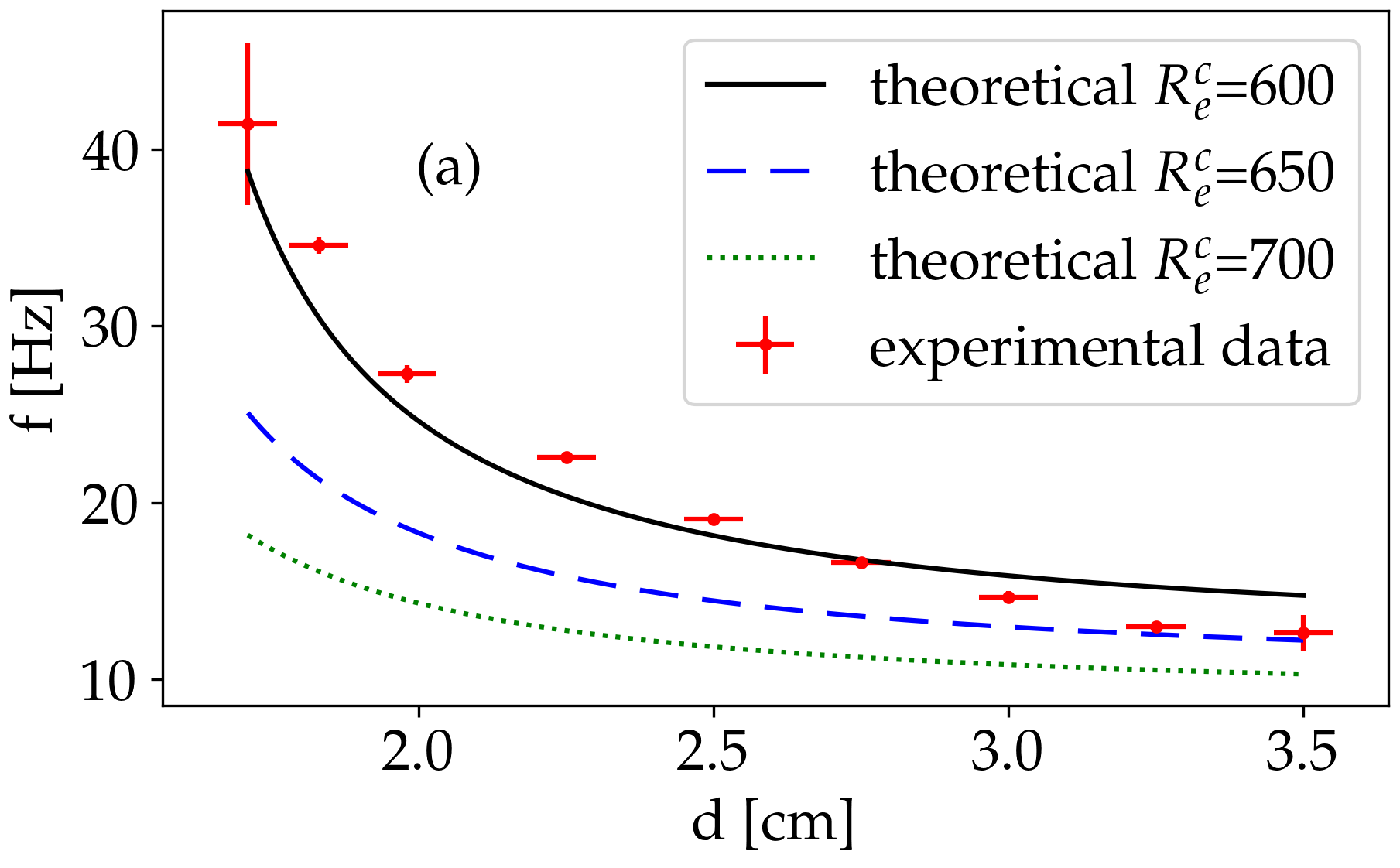}
         \includegraphics[width=0.45\textwidth]{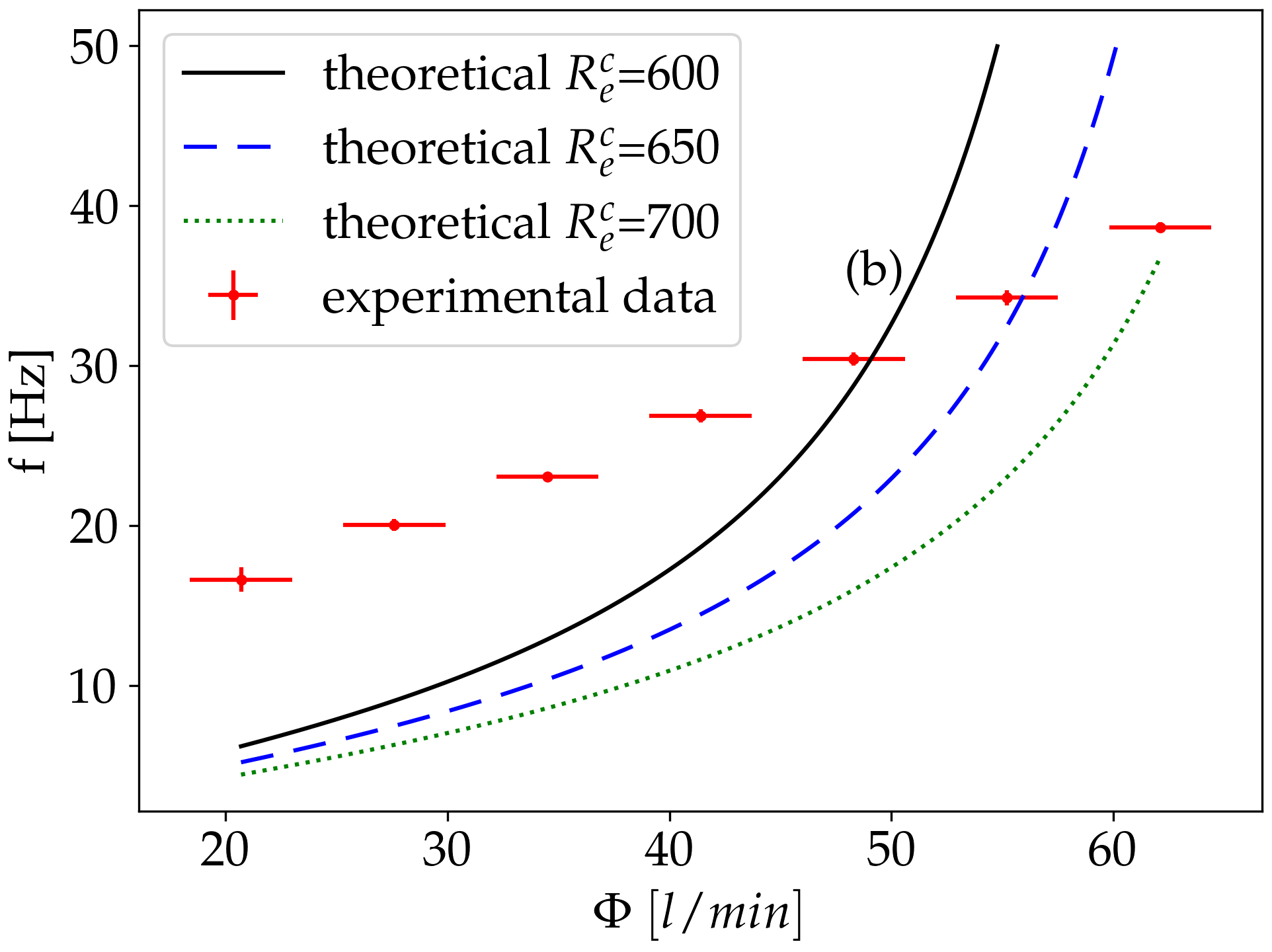}
        \caption{
The Helium column's oscillation frequency predicted by approximation (\ref{eq:rey3}) and the results observed in the experiments. Figure (a) presents the results for different nozzle diameters and Figure (b) is the result for different outflow rate. We illustrate the theoretically predicted trend for several $R_e^c$  values as indicated in the legend.   
For the theoretical calculations we used the parameters for our experiments with Helium: $g=10\ m/s^2,\ \rho_{Air}=1.2\ kg/m^3,\ \rho_{He}=0.17\ kg/m^3,\ \nu=11.52\cdot 10^{-5}\ m^2/s$. We have chosen $\Phi=46\pm2.3\ l/min$ for Figure (a) and $r_0=0.01\ m$  for Figure (b). The model correctly predicts the experimentally obtained trends and the magnitude of the obtained frequencies.}         
\label{fig:toyModellNoFriction}
\end{figure}

Equations (\ref{eq:rey2}) and (\ref{eq:rey3}) allow now to plot the trends for the estimated oscillation frequency as a function of nozzle radius ($r_0$) and flow debit ($\Phi$).  Considering some realistic critical Reynolds number $R_e^c$ values in Figures 
\ref{fig:toyModellNoFriction}a and \ref{fig:toyModellNoFriction}b we compare the theoretical trends with the results of the experiments. The trends offered by our simple model are in good agreement with the observations and also the predicted oscillation frequencies are of the right orders of magnitude. 

\subsection{Dynamics with friction}

We reconsider now the previous approach by introducing viscosity. First we will use no-slip condition and then semi-slip boundary condition for the Helium-air interface.

\subsubsection{No-slip boundary condition}\label{ch:subsection522}

We consider now the effect of friction with no-slip boundary condition at the interface of Helium and air. In deriving the average velocity of a fluid element as a function of time, we assume that the velocity of the Helium column has a  cylindrical symmetry and the velocity $v_z(r,t)$ as a function of radius has a parabolic profile.

The equation of motion is written for the average velocity of a cylindrical volume element of height $\langle v_z(t)\rangle \cdot da$. The mass of the volume element can be given as
\begin{equation}
	\begin{split}
		dm=\rho_{He}\cdot \langle v_z\rangle \cdot\pi\cdot{r_z}^2\cdot da,
	\end{split}
	\label{eq:fric1}
\end{equation}
where $r_z$ denotes the radius of the considered element at a height $z$.
The volume element is affected by the friction force and the buoyancy force. The buoyancy force is approximated as:
\begin{equation}
	\begin{split}
		dF_{b}=(\rho_{Air}-\rho_{He})\cdot \langle v_z\rangle \cdot\pi\cdot{r_z}^2\cdot g\cdot da
	\end{split}
	\label{eq:fric2}
\end{equation}
The friction force is derived from the shear stress $\kappa$, which for a $\mu$ kinematic viscosity can be written as:
\begin{equation}
	\begin{split}
		\kappa=\frac{dv_z}{dr}\cdot \mu
	\end{split}
	\label{eq:fric3}
\end{equation}
The friction force between the air and the Helium is then:
\begin{equation}
		dF_{f}=\mu\cdot 2\cdot \pi\cdot r_z \cdot \langle v_z\rangle  \cdot da\cdot \left. \frac{dv_z}{dr}\right|_{r=r_z}
	\label{eq:fric4}
\end{equation}
For the no-slip boundary condition the velocity $v_z$ as a function of radius is described by the  parabolic profile: 
\begin{equation}
	\begin{split}
		v_z(t,r)=2\cdot \langle v_{z}(t)\rangle \cdot\left(1-\frac{r^2}{{r_z}^2}\right)
	\end{split}
	\label{eq:fric5}
\end{equation}
Using the above equation the friction force from (\ref{eq:fric4}) can be estimated as: 
\begin{equation}
	\begin{split}
		dF_{f}=-8\cdot \pi\cdot \mu \cdot {\langle v_z\rangle }^2 \cdot da
	\end{split}
	\label{eq:fric6}
\end{equation}
Assuming that the parabolic profile is maintained throughout the acceleration, the equation of motion for the chosen cylindrical element can be written using the average velocity :
\begin{equation}
	\begin{split}
		dm\frac{d\langle v_z\rangle }{dt}=dF_{f}+dF_{b}
	\end{split}
	\label{eq:fric7}
\end{equation}
Using now the formula for $dm$, buoyancy and friction forces given by equations (\ref{eq:fric1}), (\ref{eq:fric2}) and (\ref{eq:fric6}), respectively,  we get:
\begin{equation}
	\begin{split}
		\rho_{He} \langle v_z\rangle \pi\cdot{r_z}^2 \frac{d\langle v_z\rangle }{dt}=-8 \pi\cdot \mu {\langle v_z\rangle }^2+(\rho_{Air}-\rho_{He}) \langle v_z\rangle\pi  {r_z}^2  g
	\end{split}
	\label{eq:fric8}
\end{equation}
The  $r_z$ radius can be estimated from the continuity equation:
\begin{equation}
	\begin{split}
		\pi\cdot{r_z}^2\cdot \langle v_z\rangle =\Phi
	\end{split}
	\label{eq:fric9}
\end{equation}
Rearranging now equation (\ref{eq:fric8}), we are lead to

\begin{equation}
	\begin{split}
		\frac{d\langle v_z\rangle }{dt}=\frac{-8\cdot \pi\cdot \mu \cdot {\langle v_z\rangle }^2}{\rho_{He}\cdot \Phi}+\frac{(\rho_{Air}-\rho_{He})\cdot g}{\rho_{He}},
	\end{split}
	\label{eq:fric10}
\end{equation}
where we introduced the notations:
\begin{equation}
	\begin{split}
		A=\frac{8\cdot \pi\cdot \mu}{\rho_{He}\cdot \Phi}\\
		B=\frac{(\rho_{Air}-\rho_{He})\cdot g}{\rho_{He}}
	\end{split}
	\label{eq:fric11}
\end{equation}
Separating the variables  $t$ and $\langle v_z\rangle $ and integrating  equation (\ref{eq:fric10})
\begin{equation}
	\begin{split}
		\int_{v_z(0)}^{v_z(t)}\frac{d\langle v_z\rangle }{B-A\cdot {\langle v_z\rangle }^{2}}=\int_{0}^{t}dt,
	\end{split}
	\label{eq:fric12}
\end{equation}
finally leads to:
\begin{equation}
	\begin{split}
		\frac{\tanh ^{-1}\left(\sqrt{\frac{A}{B}}\cdot \langle v_z(t)\rangle \right)-\tanh ^{-1}\left(\sqrt{\frac{A}{B}}\cdot \langle v_z(0)\rangle \right)}{\sqrt{A\cdot B}}=t
	\end{split}
	\label{eq:fric13}
\end{equation}
In order to estimate the oscillation frequency the average velocity $\langle v_z(t)\rangle $ is expressed from the Reynolds number and $\langle v_z(0)\rangle $ is expressed from the continuity equation as a function of flow yield $\Phi$ and nozzle radius $r_0$.
Using equations (\ref{eq:ray}) and (\ref{eq:fric9}), the velocity can be given as:
\begin{equation}
	\begin{split}
		\langle v_z\rangle =\frac{\pi\cdot{R_e}^{2}\cdot\nu^2}{4\cdot\Phi}
	\end{split}
	\label{eq:fric14}
\end{equation}
From equations (\ref{eq:fric9}), (\ref{eq:fric12}) and (\ref{eq:fric13}) the oscillation frequency as a function of the critical Reynolds number is derived:
\begin{equation}
	\begin{split}
		f=\frac{\sqrt{A\cdot B}}{\tanh ^{-1}\left(\sqrt{\frac{A}{B}}\cdot \frac{\pi\cdot{R_{e}^c}^{2}\cdot\nu^2}{4\cdot\Phi}\right)-\tanh ^{-1}\left(\sqrt{\frac{A}{B}}\cdot \frac{\Phi}{\pi\cdot{r_0}^2}\right)}
	\end{split}
	\label{eq:fric15}
\end{equation}

Let us examine now the predictions of the model with the implemented no-slip boundary conditions and friction. First we conclude that it is not possible to find a critical Reynolds number for which the above model gives a positive oscillation frequency for the entire experimentally studied flow yield range. The reason for this is the unrealistic no-slip boundary condition. In reality one would expect that the Helium column 
induces a strong flow in the surrounding air. This airflow depends on the Helium yield, therefore a correct frequency formula would require a correction term that depends on the flow yield.
Since we do not have experimental data on the dependence of the airflow rate on the Helium flow yield, we cannot estimate this term. Assuming however that this correction term depends only on the yield, equation (\ref{eq:fric15}) could give a good approximation for a constant yield.
In the experimental study where the effect of the nozzle diameter was investigated, the Helium flow rate was fixed, therefore accepting the above argument, our toy-model could approximate the experimental results well in this case.

\begin{figure}
\centering 
\includegraphics[width=0.45\textwidth]{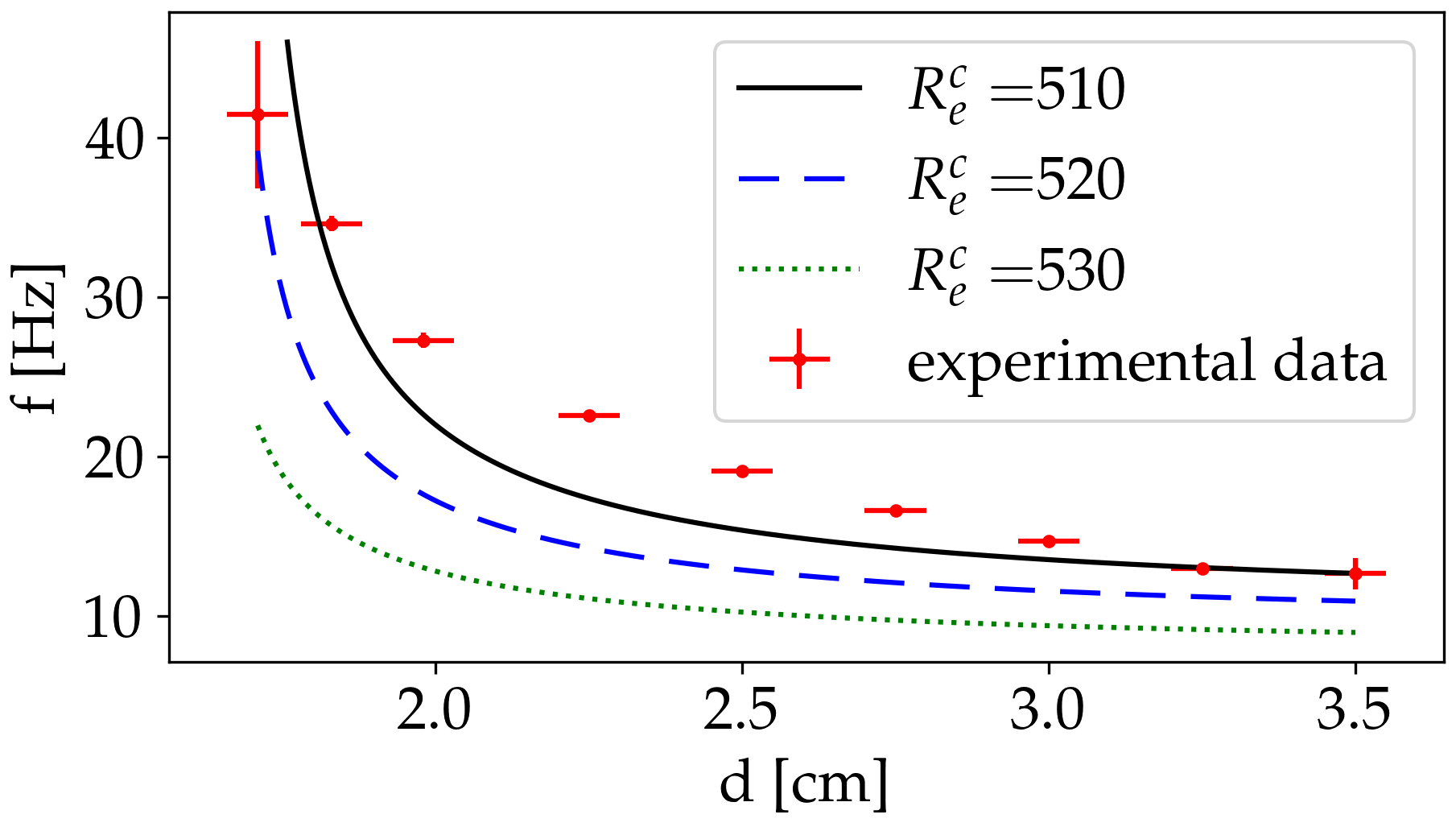}
\caption{Oscillation frequencies predicted by approximation (\ref{eq:fric15}) (continuous line) with different $R_e^c$ choices in comparison with the experimentally observed oscillation frequency as a function of the nozzle diameter. The following parameters were used in the calculations: $g=10\ m/s^2,\ \rho_{Air}=1.2\ kg/m^3,\ \rho_{He}=0.17\ kg/m^3,\ \nu=11.52\cdot 10^{-5}\ m^2/s,\ \mu=1.96\cdot10^{-5}\ Pa\cdot s,\ \Phi=46\pm 2.3\ l/min$. }
\label{fig:modellfriction}
\end{figure}

In Figure \ref{fig:modellfriction}, the experimentally measured oscillation frequency as a function of the nozzle diameter is compared with the theoretical prediction for no-slip boundary condition using different $R_e^c$ values, as indicated in the legend. 
For the model we used the following parameters  : 
$g=10\ m/s^2,\ \rho_{Air}=1.2\ kg/m^3,\ \rho_{He}=0.17\ kg/m^3,\ \nu=11.52\cdot 10^{-5}\ m^2/s,\ \mu=1.96\cdot10^{-5}\ Pa\cdot s,\ \Phi=46\pm 2.3\ l/min,\ R_{e}^c=519$.

\subsubsection{Semi-slip boundary condition}\label{ch:subsection522}%

Let us briefly discuss how the application of a non-zero velocity for the Helium-air boundary layer would improve the predictions of the model.
Since the derivation steps are almost identical to those presented in the previous subsection, only the differences are outlined in the followings.
We assume again that the velocity profile of the Helium column can be described by a parabolic function as a function of the radius $r$, the only difference from the no-slip boundary condition is that now the velocity at $r_z$  should not be set to 0. 
Based on the above consideration, the radial profile of the speed can be approximated in the form
\begin{equation}
		v_z(t,r)=\frac{2\cdot \langle v_{z}(t)\rangle} {\left(2-\beta\right)} \cdot \left (1-\frac{r^2}{{r_z}^2}\cdot \beta \right ),
	\label{eq:fric25}
\end{equation}
where $\beta$ can take a value between $0$ and $1$. This constant governs the difference of the velocity at $r=r_z$ relative to the value $0$,  considered for no-slip boundary conditions. For $\beta=0$ we get the frictionless case while for $\beta=1$ we get the no-slip boundary condition. In between this extreme values is reality. 

\begin{figure}
         \includegraphics[width=0.45\textwidth]{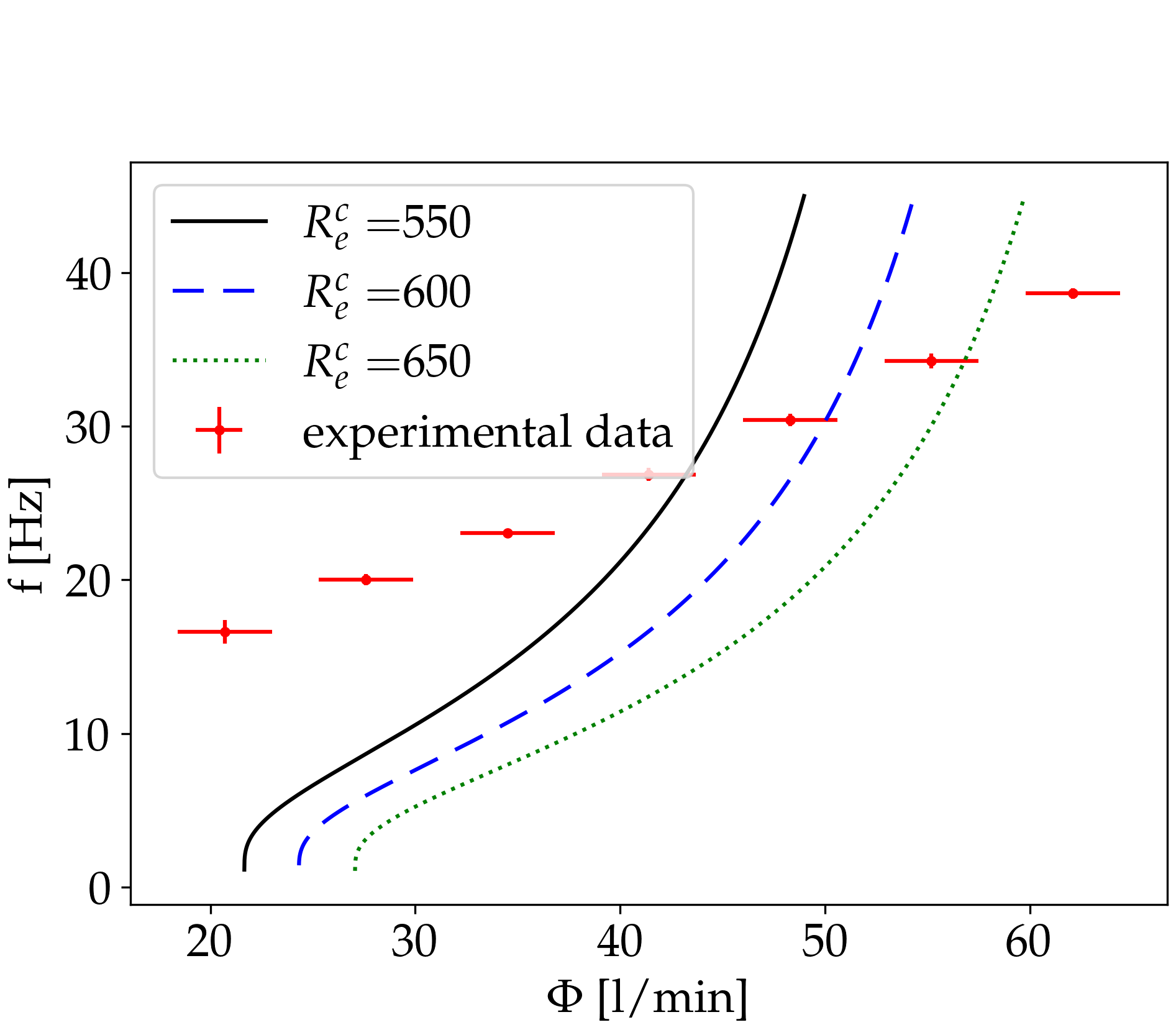}
         \includegraphics[width=0.45\textwidth]{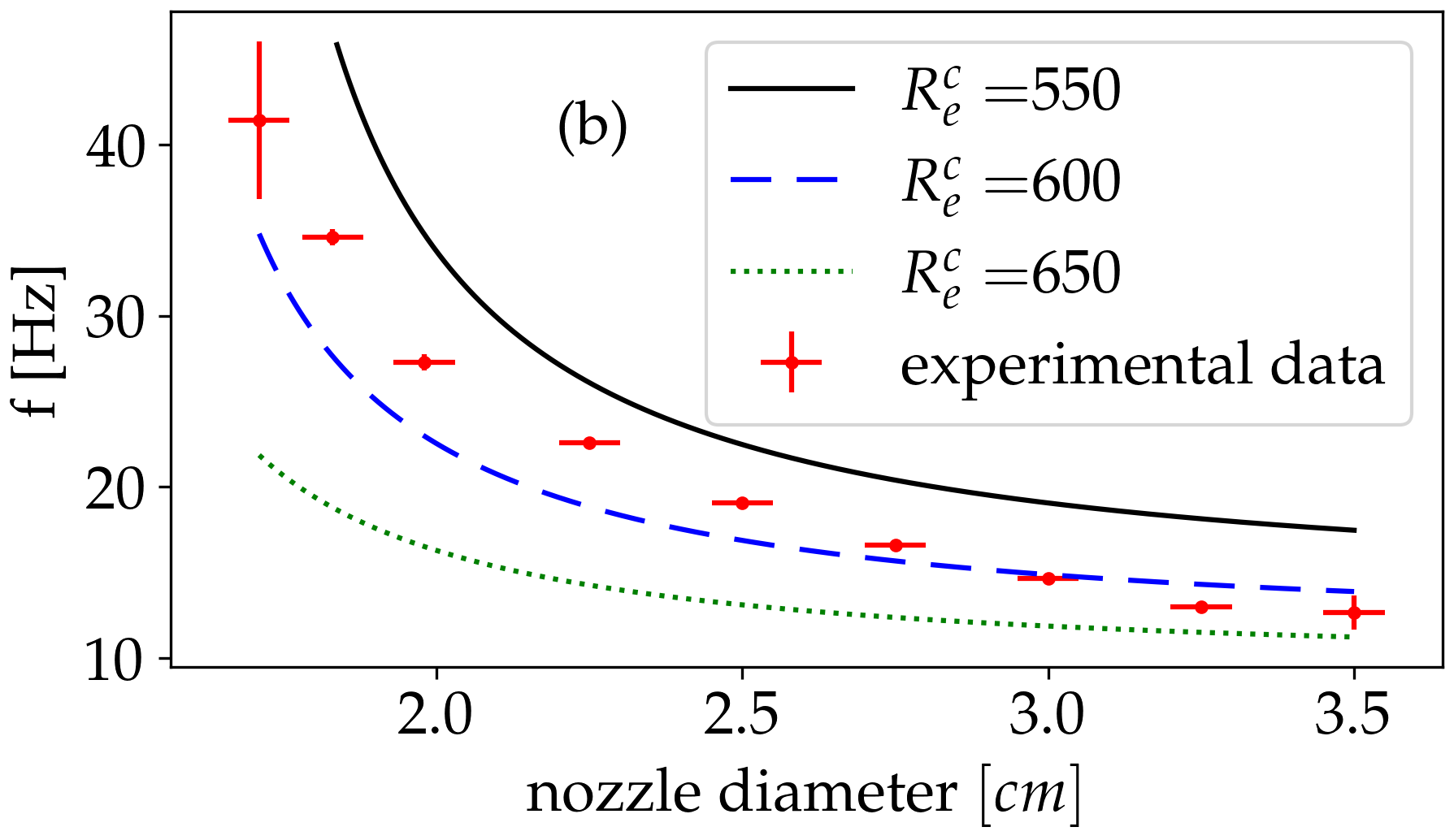}
        \caption{Experimentally observed oscillation frequencies for the Helium column oscillations in comparison with the frequency values predicted by the equation (\ref{eq:fric15})  particularised  for semi-slip boundary conditions using $\beta=0.18$ and several reasonable $R_e^c$ values as indicated in the legend.  Figure (a) is for the oscillation frequency as a function of the flow yield (nozzle diameter is $r_0=2cm$) and Figure (b) shows the frequency as a function of the nozzle diameter for a yield of $\Phi=46$ l/min. In agreement with our experimental setup the following parameters were used: $g=10\ m/s^2,\ \rho_{Air}=1.2\ kg/m^3,\ \rho_{He}=0.17\ kg/m^3,\ \nu=11.52\cdot 10^{-5}\ m^2/s,\ \mu=1.96\cdot10^{-5}\ Pa\cdot s$}
        \label{fig:toysemislip}
\end{figure}

The shape of the velocity profile is contained only in the equation of the friction force. Using this form in  equation (\ref{eq:fric4}), the friction force becomes:
\begin{equation}
	\begin{split}
		dF_{f}=-8\cdot \pi\cdot \mu \cdot {\langle v_z\rangle }^2\cdot \frac{\beta}{2-\beta} \cdot da
	\end{split}
	\label{eq:fric26}
\end{equation}
From this point on we continue with the same straightforward steps that are used for the no-slip boundary condition leading to a difference in the value of $A$:
\begin{equation}
	\begin{split}
		A=\frac{8\cdot \pi\cdot \mu\cdot\beta}{\rho_{He}\cdot \Phi\cdot(2-\beta)}
	\end{split}
	\label{eq:fric11}
\end{equation}
Using the $A$  given above we are lead to a model with two free parameters, $\beta$ and $R_e^c$.  
 Considering now the following experimental parameters:  $g=10\ m/s^2,\ \rho_{Air}=1.2\ kg/m^3,\ \rho_{He}=0.17\ kg/m^3,\ \nu=11.52\cdot 10^{-5}\ m^2/s,\ \mu=1.96\cdot10^{-5}\ Pa\cdot s$ and $\beta=0.18$ we find an acceptable trend for the oscillation frequencies as a function of the nozzle diameter and flow yield. Results for some reasonable $R_e^c$ values are summarized in Figure \ref{fig:toysemislip}.

\section{Discussion and Conclusions}

Oscillation of diffusion flames is a complex phenomenon in which both, physical and chemical processes concomitantly takes place. Consequently, it is difficult to conclude which are the relevant processes responsible for this behavior. In our previous work \cite{Gergely2020}, we explained the oscillation of candle flames and their synchronization using a dynamical model that incorporates the effects of the chemical reaction taking place during the combustion. Although the proposed model described the collective behavior and the experimentally obtained trends for the oscillation frequency well, there are experimental results suggesting that similar oscillations can be explained solely by hydrodynamic instabilities \cite{Yuan1994}. In the present work we used both experimental and theoretical approaches to investigate whether the hydrodynamic instabilities that occur in rising gas columns can explain the relevant features of the oscillations observed for the diffusion flames.

First, we have shown experimentally that for certain experimental parameters rising  gas columns can induce oscillations that are similar to the one observed in diffusion flames. The quantitative analysis of the oscillation in convective flow was performed for a rising Helium gas column. For such a Helium flow, we have shown that for a constant flow-rate the frequency of the oscillation decreases with increasing nozzle diameter, similarly with the flickering frequency of candle bundles as a function of the number of candles in the pack. The decreasing trend observed for the Helium flow can be well approximated by a power function with $-1.64$ power exponent.

For a constant nozzle diameter, the oscillation frequency of Helium column increases roughly linearly with the floe debit of Helium. For two interacting Helium columns a collective oscillation similar with the one observed for the diffusion flames were observed.

The main difference however relative to diffusion flames is that for the Helium columns only counter-phase synchronization is observable.  It is thus conceivable that the mechanism leading to synchronization is fundamentally different but it is also possible that the interaction between the Helium columns was not strong enough to find the in-phase synchronization at short separation distances. Another possible explanation could be that for diffusion flames whenever one observes the in-phase synchrony the involved hydrodynamic flows 
are not really separated but visually we detect the hot part of the flames as separate regions.
When collective behavior was found, we examined the dependence of the oscillation frequency as a function of the separation distance. Here we observed a similar decreasing trend as the one observed for the counter-phase synchronization in diffusion flames.

The observed oscillation with slightly different frequencies for two interacting Helium columns is similar to the phenomenon of beating, known in acoustics. This was observed in the case of diffusion flames as well and it was reported by Chen et. al \cite{Chen2019}. In our experiments beating was observed both by using different nozzle diameters at the same flow yield and with different yields and the same nozzle diameter of the two Helium columns.

In order to approach the observed oscillations theoretically we proposed a simplified but analytically treatable model. Our main assumption is built on the observation that the Reynolds number of the Helium column that is accelerating under the effect buoyancy increases with time until reaching a critical Reynolds number where the flow becomes unstable. The oscillation frequency was approximated as the time needed to reach this situation.  The theoretical model was discussed in two cases: a frictionless approach and another approach with taking into account friction as well.  The simple frictionless case offered already an unexpectedly good trend for the oscillation frequency as a function of the nozzle diameter and the outflow yield. The model discussed by incorporating friction with no-slip boundary condition at the Helium-air interface gives a better agreement for the dependence of the experimentally measured oscillation frequency as a function of nozzle size but it fails in describing the experimental results observed for the oscillation frequency as a function of the flow rate.  Using instead of the no-slip boundary condition semi-slip boundary conditions the model results are improved at the cost of an extra fitting parameter. We are confident that by taking into account also the extra resistance of the flow due to the developed instability, the predictions of such an approach can be significantly improved. Unfortunately our theoretical approach was not suitable to account for the collective behavior of the oscillating flows.

In conclusion, we can state that our study proves our hypothesis according to which the instabilities in a Helium column ascending from a circular nozzle into the air behaves in a similar manner to the oscillation of diffusion flames. Seemingly the hydrodynamic processes by their own are able to explain the oscillations observed for the diffusion flames. The extremely simple analytical approach considered by us for this complex phenomenon, leads to qualitatively good trends and right orders of magnitude for the oscillation frequency.  This promising agreement suggests once again the power of simple, analytically solvable models in approaching universalities in complex systems.

\vspace{2cm}
{\bf Acknowledgment} Work supported by the UEFISCDI PN-III-P4-ID-PCE-2020-0647 research grant. The work of A. Gergely is also supported by the Collegium Talentum Programme of Hungary

\end{document}